\documentclass[a4paper,12pt]{article}
\usepackage{amsmath}
\usepackage{graphicx}

\begin{document}

\begin{center}

{\Large \bf Bulk Higgs and Gauge fields in a multiply warped braneworld model}\\[20mm]

Ashmita Das\footnote{E-mail address: tpad@iacs.res.in},
R. S. Hundi\footnote{E-mail address: tprsh@iacs.res.in} and
Soumitra SenGupta\footnote{E-mail address: tpssg@iacs.res.in}\\
Department of Theoretical Physics,\\
Indian Association for the Cultivation of Science,\\
2A $\&$ 2B Raja S.C. Mullick Road,\\
Kolkata - 700 032, India.\\[20mm]

\end{center}

\begin{abstract}
We readdress the problems associated with bulk Higgs and the gauge fields in  
a 5-dimensional Randall-Sundrum model by extending the model to
six dimensions with double warping along the two extra spatial dimensions. In this
6-dimensional model we have a freedom of two moduli scales as against one
modulus in the 5-dimensional model. With a little hierarchy between these moduli
we can obtain the right magnitude for  $W$ and $Z$ boson
masses from the Kaluza-Klein  modes of  massive bulk gauge fields where the spontaneous symmetry breaking is triggered by 
bulk Higgs . We also have determined the gauge couplings of the standard model fermions with
Kaluza-Klein modes of the gauge fields. Unlike the case of 5-dimensional model with a massless bulk gauge field,
here we have shown that the gauge couplings and the
masses of the Kaluza-Klein gauge fields satisfy the precision electroweak constraints
and also obey the Tevatron bounds. 

\end{abstract}

\newpage

\section{Introduction}
\label{S:intro}
The hierarchy between electroweak and Planck scales can be
addressed in extra dimensional models. Among these, the model proposed
by Randall and Sundrum (RS) assumes warp geometry of the
space-time in 5 dimensions \cite{RS}. The fifth dimension has Planck
scale length $r_c$ and is compactified on the space $S^1/Z_2$.
Two 3-branes are supported on either side of this fifth
dimension. The exponential suppression
along the fifth dimension naturally suppresses Planck scale
quantities of one 3-brane into electroweak scale on the second
3-brane, which is identified as TeV-brane and can be interpreted
as our universe. In the original RS model the standard model
fields are assumed to lie on the TeV-brane while only gravity propagates in
the bulk. Later works have explored the phenomenology of bulk standard model fields
in warped geometry model \cite{GW,DHR,Pom,GN,CHNOY}. It was however shown that
the models with bulk gauge and Higgs fields where the spontaneous symmetry breaking
takes place in the bulk, encounters serious problems. 
The two main problems are:\\

(i) The non-Abelian gauge fields acquire masses through the Higgs vacuum
expectation value (vev) generated through spontaneous symmetry breaking 
in the bulk. This vev being a bulk parameter has a magnitude of the order
of the Planck scale and therefore lends a very large bulk mass 
$\sim$ Planck mass to the gauge boson in the bulk. As a result the lowest lying masses
in the KK tower of the gauge boson 
on the visible brane becomes $\sim$ TeV which fails to comply with the $W$ and
$Z$ boson masses $\leq 100$ GeV. If one 
tries to reduce it by adjusting the bulk parameters then that would jeopardize
the unique feature of Planck scale to TeV scale warping,
{\it i.e.} the resolution of gauge hierarchy problem which was the original
motivation of such a warped geometry model.

(ii) For an Abelian gauge boson with zero bulk mass, the massless Kaluza-Klein (KK)
mode on the TeV-brane corresponds to
photon. However, the first excited state in the KK tower has an unacceptably
large coupling with fermions.  This puts a stringent bound on the mass of this
state such that the model may survive the direct search bound at the Fermilab 
Tevatron as well as  precision electroweak constraints. However, the mass of this
first KK mode turns out to be much lower than the above bound. Once again it is
impossible either to reduce the coupling or to increase the mass 
by adjusting the bulk parameters without disturbing the resolution of the gauge hierarchy/fine tuning problem.\\

We refer our readers to \cite{DHR, CHNOY,DMS,higgs-others2,ssg1,cust} where both
these problems have been discussed in details. 
Recently, in a generalized 5-dimensional RS model with non-flat visible brane, by adjusting
the brane cosmological constant the problem coming from the
precision electroweak tests have been averted and also the bulk
Higgs problem has been resolved \cite{DMS}. 
The  brane cosmological constant however was found to be negative implying that the visible brane in such a case is an 
anti-de Sitter 3-brane.
 
In the present work
we address these problems from a different viewpoint {\it i.e.} in the backdrop of a 6-dimensional doubly
warped model with flat 3-branes 
which is an extension of the original RS model to more than one extra dimensions\cite{CS}. In this 6-dimensional
model two extra spatial coordinates are compactified such that
the space-time manifold is $[M^{(1,3)}\times S^1/Z_2]
\times S^1/Z_2$. Compared to the RS model, here the two
extra dimensions, denoted by angular coordinates $y,z$,
are doubly warped. Four 4-branes are located at the orbifolded
points: $y=0,\pi$, $z=0,\pi$. The intersection of any two
4-branes gives a 3-brane. The 3-brane located at $(y,z)=(\pi,0)$
is identified with our universe. Analogous to the RS setup,
the mass scale suppression can be felt along both the coordinates
$y,z$. We can choose the moduli of these coordinates, say
$R_y$ and $r_z$, such that TeV scale masses can be
generated on the visible brane located at $(y,z)=(\pi,0)$. Since
there is an extra freedom through an additional modulus in this model compared to the 5-dimensional RS model,
we explore if the above problems relating to bulk Higgs can be solved
in the 6-dimensional doubly warped model by adjusting the
moduli $R_y,r_z$ suitably.

We organize our paper as follows.
In the following  section we explain some essential features of
the 6-dimensional doubly warped model. In Sec. 3 we
describe the KK mode analysis of 
gauge bosons and fermions in 6-dimensional bulk and the corresponding modes on the visible 3-brane. In
Sec. 4 we present our results and argue that the precision
electroweak tests put no  additional constraints on this
model. We conclude in Sec. 5.

\section{The 6-dimensional doubly warped model}
\label{S:mod}

As explained previously, the 6-dimensional doubly warped
model has space-time of six dimensions and the extra two
spatial dimensions are orbifolded by $Z_2$ symmetry \cite{CS}. The
manifold under consideration is $[M^{(1,3)}\times S^1/Z_2]
\times S^1/Z_2$ with four non-compact dimensions denoted
by $x^\mu$, $\mu=0,\cdots,3$. Since we are interested in
doubly warped model, the metric in this model can be chosen
as
\begin{equation}
ds^2 = b^2(z)[a^2(y)\eta_{\mu\nu}dx^\mu dx^\nu +R_y^2dy^2]+r_z^2dz^2
\label{E:metric}
\end{equation}
As explained before the angular coordinates $y,z$ represent the
extra spatial dimensions with moduli $R_y,r_z$, respectively. The
Minkowski matrix in the usual 4-dimensions has the form $\eta_{\mu\nu} = {\rm diag}
(-1,1,1,1)$. The functions $a(y),b(z)$ give warp factors in
the $y$ and $z$ directions, respectively. The total bulk-brane action
of this model has a form \cite{CS}
\begin{eqnarray}
S &=& S_6+S_5
\nonumber \\
S_6 &=& \int d^4xdydz\sqrt{-g_6}(R_6-\Lambda), \quad
\nonumber \\
S_5 &=& \int d^4xdydz[V_1\delta(y)+V_2\delta(y-\pi)]
     +\int d^4xdydz[V_3\delta(z)+V_4\delta(z-\pi)]
\end{eqnarray}
Here, $V_{1,2}$ and $V_{3,4}$ are brane tensions of the branes located at
$y=0,\pi$ and $z=0,\pi$, respectively. $\Lambda$ is the cosmological
constant in 6-dimensions. The 3-branes are located at the intersection points of the four 4-branes.

After solving Einstein's
equations the solutions to the warp functions
of the metric as given in eq. (\ref{E:metric}) have a form \cite{CS}
\begin{eqnarray}
a(y) &=& \exp(-c|y|), \quad b(z) = \frac{\cosh(kz)}{\cosh(k\pi)}
\nonumber \\
c&\equiv & \frac{R_yk}{r_z\cosh(k\pi)},\quad k\equiv r_z\sqrt{
\frac{-\Lambda}{10M_P^4}}
\label{E:sol}
\end{eqnarray}
Here, $M_P$ is the Planck scale.
The warp factors $a(y)$ and $b(z)$ give
largest suppression from $y=0, z=\pi$ brane to $y=\pi z=0$ brane. For this reason
we can interpret the 3-brane formed out of the intersection of
4-branes at $y=\pi$ and $z=0$ as our standard model brane. The suppression
factor $f$ on the standard model brane can be written as
\begin{equation}
f =\frac{\exp(-c\pi)}{\cosh(k\pi)}
\label{E:supp}
\end{equation}
The desired suppression of $10^{-16}$ on the standard model brane
can be obtained for different combinations of the parameters $c$ and $k$. However, from
the relation for $c$ in eq. (\ref{E:sol}) it can be noticed that
in order not to have large hierarchy in the moduli $R_y$ and $r_z$, either of $c$
or $k$ must be large where as other is small, e.g. $c\sim$ 11.4 and $k\sim$ 0.1.
This implies that the warping along $y$ is large whereas that along $z$ is small.
It has been argued that this feature may offer an explanation of the small mass hierarchy
among the standard model fermions\cite{CS}.  

The 6-dimensional model which has been described here is thus viable
in explaining the hierarchy between Planck scale and the electroweak
scale without introducing large hierarchy between the moduli
$R_y$ and $r_z$. In this model KK modes of bulk scalar fields have
been studied \cite{KMS1}. Bulk fermion fields have also been
studied in this model with a possibility of localizing
them on a 4-brane \cite{KMS2}. However the possibility of bulk gauge and Higgs field in this
model has not been explored yet. In the following section we derive the KK modes of the gauge field, fermion fields  and the corresponding
couplings to estimate the viability of this model in respect to the problems discussed earlier. 
We reiterate that our aim is to explore if we can put the Higgs in the bulk of such 6-dimensional
multiply warped model without invoking any contradiction with the precision electroweak test 
\cite{DHR,CHNOY} as was encountered in the 5-dimensional RS model.
\section{Gauge bosons and fermions in the bulk}

In this section we explain the KK decomposition and eigenvalue
equations of KK gauge bosons
and KK fermions which arise from the respective bulk fields after
integrating over the two extra dimensions of the model.

\subsection{KK modes of the gauge bosons}
For simplicity, we consider a U(1) gauge theory, but our derivation
given below is applicable to non-Abelian theory as well. In a realistic
model the gauge fields can acquire non-zero masses due to spontaneous
symmetry breaking. In our model, Higgs mechanism can take place in the
bulk of the 6-dimensions and the vev of the Higgs
field will be of the order of Planck scale. The vev of the Higgs field
contributes to generate the bulk mass for the gauge field. Hence, after spontaneous
symmetry breaking the invariant action can be written as
\begin{equation}
S_G = \int d^4xdydz\sqrt{-G}\left(-\frac{1}{4}G^{MK}G^{NL}F_{KL}F_{MN}
-\frac{1}{2}M^2G^{MK}A_MA_K\right),
\label{E:SG}
\end{equation}
where, $M$ is the bulk mass $\sim M_P$ and $G={\rm det}(G_{AB})$ is the determinant
of the metric $G_{AB}$ which is given in eq. (\ref{E:metric}).
$F_{KL}=\partial_K A_L-\partial_LA_K$ is the gauge field strength. Exploiting the
gauge symmetry we can choose the gauge where
$A_4=A_5=0$. The KK decomposition of the gauge field can be taken as
\begin{equation}
A_\mu = \sum_{n,p}A^{(n,p)}_\mu(x)\xi_n(y)\chi_p(z)/\sqrt{R_yr_z}.
\label{E:KKgau}
\end{equation}
The KK fields in the 4-dimensions $A_\mu^{(n,p)}$ carry two indices
$n,p$ due to the two additional dimensions of the model. The functions
$\xi_n(y)$ and $\chi_p(z)$ give KK wave functions in the $y$ and $z$ directions,
respectively. Substituting the above KK decomposition in eq. (\ref{E:SG})
and integrating over the $y$ and $z$ coordinates, we demand
that the resulting action in the 4-dimensions must have a form
\begin{equation}
\sum_{n,p}-\frac{1}{4}F_{\mu\nu}^{(n,p)}F^{(n,p)\mu\nu}-\frac{1}{2}
m_{n,p}^2A_{\mu}^{(n,p)}A^{(n,p)\mu},
\end{equation}
where $m_{n,p}$ is the mass of the KK field $A_\mu^{(n,p)}$. This can be achieved provided the KK wave functions 
satisfy the following orthonormality condition.
\begin{equation}
\int dy~\xi_n(y)\xi_{n^\prime}(y) = \delta_{nn^\prime},
\quad
\int dz~b(z)\chi_p(z)\chi_{p^\prime}(z) = \delta_{pp^\prime}.
\label{E:gnorm}
\end{equation}
Moreover, in addition to the above normalization conditions the following
eigenvalue equations for the $\xi_n$ and $\chi_p$ must also be satisfied:
\begin{eqnarray}
\frac{1}{R_y^2}\partial_y(a^2\partial_y\xi_n)-m_p^2a^2\xi_n &=& -m_{n,p}^2\xi_n,
\nonumber \\
\frac{1}{r_z^2}\partial_z(b^3\partial_z\chi_p)-M^2b^3\chi_p &=& -m_{p}^2\chi_p.
\label{E:KKyz}
\end{eqnarray}
Here, $m_p$ is a mass parameter which is determined by solving the equation
for $\chi_p(z)$, and the value of $m_p$ determines the KK mass $m_{n,p}$ through
the eigenvalue equation for $\xi_n$, as given above.

The second of the eq. (\ref{E:KKyz}) can be solved by approximating
$b(z)\sim\exp[-k(\pi-z)]=\exp[-k\tilde{z}]$. By writing
$\tilde{\chi}_p(z)=\exp(-3k\tilde{z}/2)\chi_p(z)$ the eigenvalue
equation for $\tilde{\chi}_p$ takes the form,
\begin{equation}
z_p^2\frac{d^2\tilde{\chi}_p}{dz_p^2}+z_p\frac{d\tilde{\chi}_p}{dz_p}
+(z_p^2-\nu_p^2)\tilde{\chi}_p=0,
\end{equation}
where $z_p=\frac{m_p}{k^\prime}\exp(k\tilde{z})$ and $\nu_p^2=\frac{9}{4}
+\left(\frac{M}{k^\prime}\right)^2$. Here, $k^\prime=k/r_z$.  The
solutions to the above equation are Bessel functions of order $\nu_p$,
and we can write
\begin{equation}
\chi_p(z)=\frac{1}{N_p}\exp(\frac{3}{2}k\tilde{z})
\left[J_{\nu_p}(z_p)+b_pY_{\nu_p}(z_p)\right],
\label{E:Zeigen}
\end{equation}
where, $N_P$ and $b_p$ are some constants.
By demanding that the function $\chi_p(z)$ be continuous at the
orbifold fixed points $z=0,\pi$ we get the following approximate
solution which determines the spectrum for $m_p$.
\begin{equation}
3J_{\nu_p}(x_{\nu_p})+x_{\nu_p}(J_{\nu_p - 1}(x_{\nu_p}) - J_{\nu_p + 1}(x_{\nu_p})) = 0,
\label{E:Zeq}
\end{equation}
where $x_{\nu_p}=\frac{m_p}{k^\prime}\exp(k\pi)$. After solving
for $m_p$ using the above equation, we can compute the KK mass $m_{n,p}$
by solving the first of eq. (\ref{E:KKyz}). By writing $\tilde{\xi}_n
=\exp(-c|y|)\xi_n$, the eigenvalue equation for $\tilde{\xi}_n(y)$ becomes,
\begin{equation}
y_n^2\frac{d^2\tilde{\xi}_n}{dy_n^2}+y_n\frac{d\tilde{\xi}_n}{dy_n}
+(y_n^2-\nu_n^2)\tilde{\xi}_n=0,
\end{equation}
where $y_n=\frac{m_{n,p}}{k^\prime}\exp(c|y|)\cosh(k\pi)$ and
$\nu_n^2=1+\left(\frac{m_p}{k^\prime}\right)^2\cosh^2(k\pi)$. The solution
for $\xi_n(y)$ can be written in terms of Bessel function of order
$\nu_n$ multiplied by growing exponential factor as
\begin{equation}
\xi_n(y)=\frac{1}{N_n}\exp(c|y|)
\left[J_{\nu_n}(y_n)+b_nY_{\nu_n}(z_n)\right],
\label{E:Yeigen}
\end{equation}
where, $N_n$ and $b_n$ are some constants. Again, by demanding that
the function $\xi_n(y)$ be continuous at the orbifold fixed points
$y=0,\pi$ the following equation determines the KK
mass $m_{n,p}$.
\begin{equation}
J_{\nu_n}(x_{\nu_n})+x_{\nu_n}(J_{\nu_n - 1}(x_{\nu_n}) - J_{\nu_n + 1}(x_{\nu_n}))/2 = 0,
\label{E:Yeq}
\end{equation}
where,
\begin{equation}
x_{\nu_n}=\frac{m_{n,p}}{k^\prime}\exp(c\pi)\cosh(k\pi).
\label{E:xn}
\end{equation}

The actual KK mass of a gauge field is found by first solving the
eq. (\ref{E:Zeq}) for $m_p$ and then solving the eq. (\ref{E:Yeq}), which
is described in the previous paragraph. The wave function of these
KK gauge fields is product of wave functions given in eqs. (\ref{E:Zeigen})
and (\ref{E:Yeigen}). \\
In the above analysis if we put the bulk gauge boson mass $M = 0$, we easily
obtain the various KK mode solutions and the corresponding masses. In this case
the lowest lying KK mode is massless which corresponds to the standard model photon.

A nice feature of the KK gauge fields in the
6-dimensional doubly warped model is that their wave functions
can be decomposed into product of functions in the two extra dimensions,
a feature which may not be evident for the bulk fermion fields which
is the subject of the next subsection.

\subsection{KK modes of the fermions}

The invariant action for a bulk fermion field $\Psi$ in 6-dimensions is
\cite{DHR,GN,CHNOY}
\begin{equation}
S_f = \int d^4xdydz\sqrt{-G}\left\{E^A_a\left[\frac{i}{2}\left(\bar{\Psi}\Gamma^a
\partial_A\Psi - \partial_A\bar{\Psi}\Gamma^a\Psi\right)+\frac{\omega_{bcA}}{8}
\bar{\Psi}\{\Gamma^a,\sigma^{bc}\}\Psi\right] - M_f\bar{\Psi}\Psi\right\}
\label{E:Sf}
\end{equation}
Here, the capital letter $A$ denotes index in the curved space
and the lower case letters $a,b,c$ denote indices in the
tangent space.
$\omega_{bcA}$ is spin connection and $E^A_a$ is inverse vielbein.
$M_f$ is the bulk mass and $\sigma^{bc}=\frac{i}{2}[\Gamma^b,\Gamma^c]$.
The Dirac matrices $\Gamma^a$ in 6-dimensions would be 8$\times$8, and
they can be taken as \cite{BDP}:
\begin{equation}
\Gamma^\mu = \gamma^\mu\otimes\sigma^0,\quad \Gamma^4=i\gamma_5\otimes\sigma^1,
\quad \Gamma^5=i\gamma_5\otimes\sigma^2
\end{equation}
Here, $\gamma^\mu$ are the Dirac matrices in 4-dimensions and
$\gamma_5=i\gamma^0\gamma^1\gamma^2\gamma^3$. $\sigma^i$, $i=1,2,3$, are the
Pauli matrices and $\sigma^0$ is the 2$\times$2 unit matrix.
The chirality in 6-dimensions is defined by the matrix $\bar{\Gamma}=\Gamma^0
\Gamma^1\Gamma^2\Gamma^3\Gamma^4\Gamma^5$ as $\bar{\Gamma}\Psi_\pm = \pm\Psi_\pm$.
The chiral fermions in 6-dimensions have both left- and right-handed chirality
of 4-dimensions, which can be projected by the operators $P_{L,R} =
(1\mp i\Gamma^0\Gamma^1\Gamma^2\Gamma^3)/2$.

As explained in Sec. \ref{S:intro}, we are interested in estimating the gauge coupling
of standard model fermions to the KK gauge bosons. We take the bulk mass
of the fermions $M_f$ to be zero, since the masses of standard model fermions
are much below the Planck scale. The term that is associated with the spin
connection in eq. (\ref{E:Sf}) would give no contribution, since the metric
in eq. (\ref{E:metric}) is diagonal. Hence, in our particular case
of interest we expand the first term of eq. (\ref{E:Sf}),
which has the following form:
\begin{eqnarray}
S_f= &\int d^4xdydz& \left\{ b^4a^3R_yr_z i\left(
\bar{\Psi}_{+L}\Gamma^\mu\partial_\mu\Psi_{+L}+
\bar{\Psi}_{+R}\Gamma^\mu\partial_\mu\Psi_{+R}+
\bar{\Psi}_{-L}\Gamma^\mu\partial_\mu\Psi_{-L}+
\bar{\Psi}_{-R}\Gamma^\mu\partial_\mu\Psi_{-R}\right) \right.
\nonumber \\
&& \left.
+\left[\bar{\Psi}_{+L}\left(\Gamma^4 D_y+\Gamma^5 D_z\right)\Psi_{+R}
+\bar{\Psi}_{+R}\left(\Gamma^4 D_y+\Gamma^5 D_z\right)\Psi_{+L}
\right. \right.
\nonumber \\
&& \left. \left.
+\bar{\Psi}_{-L}\left(\Gamma^4 D_y+\Gamma^5 D_z\right)\Psi_{-R}
+\bar{\Psi}_{-R}\left(\Gamma^4 D_y+\Gamma^5 D_z\right)\Psi_{-L}
\right]\right\},
\label{E:Sfin}
\end{eqnarray}
where the differential operators are defined as:
$ D_y = \frac{i}{2}b^4r_z(a^4\partial_y
+\partial_ya^4)$ and $ D_z = \frac{i}{2}a^4R_y(b^5\partial_z
+\partial_zb^5)$.
In the subscript of the fields $\Psi$ the $\pm$ indicates the chirality
in 6-dimensions and the $L,R$ stands for the left- and right-handed chirality
of the 4-dimensions. Terms in line 2 and 3 of the above equation give
effective masses for the KK modes in the 4-dimensions. These terms indicate
that in general we cannot decompose the wave functions into $y$ and $z$ parts separately,
like what we have done for the KK wave functions of the gauge bosons as
described previously in eq. (\ref{E:KKgau}). Hence, for the bulk fermions the KK decomposition
can be taken as
\begin{eqnarray}
\Psi_{+L,-R}(x^\mu,y,z)&=&\frac{1}{\sqrt{R_yr_z}}\sum_{j,k}\psi^{(j,k)}_{+L,-R}(x^\mu)
f^{(j,k)}_{+L,-R}(y,z)\otimes\left(\begin{array}{c}1\\0\end{array}\right),
\nonumber \\
\Psi_{-L,+R}(x^\mu,y,z)&=&\frac{1}{\sqrt{R_yr_z}}\sum_{j,k}\psi^{(j,k)}_{-L,+R}(x^\mu)
f^{(j,k)}_{-L,+R}(y,z)\otimes\left(\begin{array}{c}0\\1\end{array}\right).
\label{E:KKfer}
\end{eqnarray}
In the above equation various fields of the form $\psi^{(j,k)}(x^\mu)$ are the KK fields
living in the 4-dimensions and $f$'s are the KK wave functions depending on both
$y$ and $z$ coordinates. Substituting the above KK decomposition into
eq. (\ref{E:Sfin}) and integrating over the $y$ and $z$ we can get the action
of the form
\begin{eqnarray}
S_f = &\int d^4x& \sum_{j,k}\bar{\psi}^{(j,k)}_{+L}i\gamma^\mu\partial\psi^{(j,k)}_{+L}
+\bar{\psi}^{(j,k)}_{+R}i\gamma^\mu\partial\psi^{(j,k)}_{+R}
+\bar{\psi}^{(j,k)}_{-L}i\gamma^\mu\partial\psi^{(j,k)}_{-L}
+\bar{\psi}^{(j,k)}_{-R}i\gamma^\mu\partial\psi^{(j,k)}_{-R}
\nonumber \\
&&
-M_{j,k}(\bar{\psi}^{(j.k)}_{+L}\psi^{(j,k)}_{+R}+\bar{\psi}^{(j.k)}_{+R}\psi^{(j,k)}_{+L}
+\bar{\psi}^{(j.k)}_{-L}\psi^{(j,k)}_{-R}+\bar{\psi}^{(j.k)}_{-R}\psi^{(j,k)}_{-L}),
\end{eqnarray}
provided the following normalization and the eigenvalue equations
for the KK wave functions are satisfied:
\begin{equation}
\int dydz b^4(z)a^3(y)\left(f^{(j,k)}_{+R,+L,-R,-L}(y,z)\right)^*
f^{(j^\prime,k^\prime)}_{+R,+L,-R,-L}(y,z) =
\delta^{j,j^\prime}\delta^{k,k^\prime},
\label{E:fnorm}
\end{equation}
\begin{eqnarray}
(i{\cal D}_y+{\cal D}_z)f^{(j,k)}_{+R}(y,z)&=&-M_{j,k}f^{(j,k)}_{+L}(y,z),
\nonumber \\
(-i{\cal D}_y+{\cal D}_z)f^{(j,k)}_{+L}(y,z)&=&-M_{j,k}f^{(j,k)}_{+R}(y,z),
\nonumber \\
(i{\cal D}_y+{\cal D}_z)f^{(j,k)}_{-L}(y,z)&=&M_{j,k}f^{(j,k)}_{-R}(y,z),
\nonumber \\
(-i{\cal D}_y+{\cal D}_z)f^{(j,k)}_{-R}(y,z)&=&M_{j,k}f^{(j,k)}_{-L}(y,z),
\label{E:feigen}
\end{eqnarray}
where the differential operators are: ${\cal D}_y=\frac{i}{2R_y}(4\partial_ya
+2a\partial_y)$ and ${\cal D}_z=\frac{i}{2r_z}a(5\partial_zb
+2b\partial_z)$.
Here, $M_{j,k}$ is the mass of the KK fermion $\psi^{(j,k)}$.

As explained previously, we are interested in standard model fermion
coupling with the KK modes of gauge field. The zero mode of the KK
fermions are identified with the standard model fermions. The wave
function for these fields can be solved from eq. (\ref{E:feigen})
by putting $M_{j,k}=0$. Here, we show the solution for the wave
function $f_{+R}^{(0,0)}(y,z)$ and the solutions for other chiral
fermions can be analogously worked out. The eigenvalue equation
we are interested in is
\begin{equation}
(i{\cal D}_y+{\cal D}_z)f_{+R}^{(0,0)}(y,z)=0,
\end{equation}
where the differential operators ${\cal D}_y$ and ${\cal D}_z$ are
defined below eq. (\ref{E:feigen}). For the zero-mode case
we can write the function $f_{+R}^{(0,0)}$ as a product of
$y$ and $z$ parts, say $f_{+R}^{(0,0)}(y,z)=f_y(y)f_z(z)$.
This simplification happens only for the zero-mode case, since
the factor $a(y)$ in the operators ${\cal D}_{y,z}$ can be taken
out and the right-hand side of the above equation is zero.
Substituting this form of $f_{+R}^{(0,0)}(y,z)$ in the above equation we get
\begin{equation}
-\frac{1}{R_y}\frac{(-4c+2\partial_y)f_y}{f_y}
+i\frac{1}{r_z}\frac{(5\partial_zb+2b\partial_z)f_z}{f_z} = 0
\end{equation}
Since the functional dependence on $y$ and $z$ are completely
separated out, we can solve for $f_y$ and $f_z$ by taking
$\frac{(-4c+2\partial_y)f_y}{f_y}=c_1$, where $c_1$ is a separation
constant. In terms of $c_1$ the functional dependences of
$f_y$ and $f_z$ are given below
\begin{equation}
f_y(y)=\exp\left(\frac{1}{2}(c_1+4c)y\right),\quad
f_z(z)=\frac{\exp\left(\frac{-ic_1r_z}{kR_y}\tan^{-1}(\tanh(kz/2))\cosh(k\pi)\right)}
{\cosh^{5/2}(kz)}
\label{E:SMwf}
\end{equation}
The value of $c_1$ can be worked out in terms of $c$ and $k$
by normalizing the wave
function $f_{+R}^{(0,0)}(y,z)$ using eq. (\ref{E:fnorm}).

\section{Bulk phenomenology}

In the previous section we have given a description of the
KK modes of the gauge bosons and fermions in the bulk of a
6-dimensional doubly warped model. Now using the mode
expansion for the bulk fields, we can calculate the gauge
couplings of standard model fermions with the KK modes of the
gauge bosons.

We now address the two problems mentioned in the beginning.
Recall that in 5-dimensional RS model it is found
that for non-zero bulk mass for non-Abelian gauge field the lowest lying
mode has mass much higher than $100$ GeV {\it i.e.} the
masses for $W$ and $Z$ bosons. Also,
for the massless gauge boson the gauge coupling with the first exited KK gauge boson
is larger than one and hence the standard model fermions are
strongly coupled \cite{DHR,CHNOY}. Due to this a stringent
lower bound of $\sim$ 10 TeV
on the mass of the first exited KK boson arose because of
the precision electroweak tests.

In this section we repeat
this exercise in the 6-dimensional doubly warped model, and
will show that due to the presence of an additional modulus we can tune
the lowest KK mode mass for non-Abelian gauge field
near $100$ GeV although the spontaneous symmetry breaking takes place in the bulk
with a bulk Higgs field with vev $\sim$ Planck scale. 
Furthermore, for a gauge boson with a zero bulk mass the coupling to mass ratio
of the first excited KK mode can survive the precision 
electroweak test without putting any  additional restriction on the model.

The action between the bulk fermions and gauge bosons
can be written as \cite{DHR,CHNOY}
\begin{equation}
S_{\rm int} = \int d^4xdydz\sqrt{-G} g_{6d}\bar{\Psi}(x^\mu,y,z)i\Gamma^aE^A_aA_A(x^\mu,y,z),
\Psi(x^\mu,y,z)
\end{equation}
where $g_{6d}$ is the gauge coupling in the 6-dimensions as has been discussed in Sec. 3.  
Substituting
the KK decomposition for the gauge and fermion fields in the above equation and also
reminding that we are working in the gauge choice where $A_4=A_5=0$,
we get the gauge coupling in the 4-dimensions as
\begin{equation}
g^{(j,k)(n,p)}_{+R} = \int dydz~g_0\pi b^4a^3\left(f_{+R}^{(j,k)}(y,z)\right)^*
f_{+R}^{(j,k)}(y,z)\xi_n(y)\chi_p(z),
\label{E:coup}
\end{equation}
where $g_0=g_{6d}/\sqrt{\pi R_y\pi r_z}$ is the effective 4-dimensional
gauge coupling. In the above equation we have given gauge coupling for
the KK fermion $\psi_{+R}^{(j,k)}$ with the KK gauge field $A_\mu^{(n,p)}$.
Similarly, the gauge couplings with the other KK fermions can be easily
obtained by replacing the $+$ with $-$ and $R$ with $L$ accordingly
in the above equation. The KK wave functions: $f$'s, $\xi$ and $\chi$,
in the above equation should be the normalized wave functions as given
by eqs. (\ref{E:gnorm}) and (\ref{E:fnorm}). However, in our particular
case of interest, where we are interested in precision electroweak tests,
we compute gauge couplings of the standard model fermions with the
KK gauge fields. Hence, the wave functions for the fermions are of
the form in eq. (\ref{E:SMwf}) and the corresponding functions for
the KK gauge fields are given in eqs. (\ref{E:Zeigen}) and (\ref{E:Yeigen}).

\begin{table}
\begin{center}
\begin{tabular}{||c|c|c||c|c|c||} \hline
\multicolumn{3}{||c||}{$\frac{M}{k^\prime}=0.5$}
& \multicolumn{3}{|c||}{$\frac{M}{k^\prime}=1.0$}\\\hline
$m_{1,2}$ = 143.78 & $m_{1,3}$ = 194.15 & $m_{1,4}$ = 244.52 &
$m_{1,2}$ = 148.33 & $m_{1,3}$ = 199.21 & $m_{1,4}$ = 249.58 \\
$\tilde{g}^{1,2}$ = 0.0015 & $\tilde{g}^{1,3}$ = 0.0028 &
$\tilde{g}^{1,4}$ = 0.0009 &
$\tilde{g}^{1,2}$ = 0.0006 & $\tilde{g}^{1,3}$ = 0.0027 &
$\tilde{g}^{1,4}$ = 0.0011\\\hline
$m_{2,1}$ = 180.23 & $m_{2,2}$ = 237.94 & $m_{2,3}$ = 295.40 &
$m_{2,1}$ = 184.53 & $m_{2,2}$ = 243.25 & $m_{2,3}$ = 300.97 \\
$\tilde{g}^{2,1}$ = 0.0127 & $\tilde{g}^{2,2}$ = 0.0012 &
$\tilde{g}^{2,3}$ = 0.0023 &
$\tilde{g}^{2,1}$ = 0.0124 & $\tilde{g}^{2,2}$ = 0.0005 &
$\tilde{g}^{2,3}$ = 0.0022 \\\hline
$m_{3,1}$ = 261.73 & $m_{3,2}$ = 322.73 & $m_{3,3}$ = 383.48 &
$m_{3,1}$ = 266.29 & $m_{3,2}$ = 328.30 & $m_{3,3}$ = 389.56 \\
$\tilde{g}^{3,1}$ = 0.0106 & $\tilde{g}^{3,2}$ = 0.0010 &
$\tilde{g}^{3,3}$ = 0.0019 &
$\tilde{g}^{3,1}$ = 0.0104 & $\tilde{g}^{3,2}$ = 0.0004 &
$\tilde{g}^{3,3}$ = 0.0019 \\\hline
\end{tabular}
\end{center}
\caption{The gauge couplings $g^{(0,0)(i,j)}$ of the standard model fermions
are given in the form $\tilde{g}^{i,j}=\frac{g^{(0,0)(i,j)}}{g_0}$, where
$g_0$ is the effective 4-dimensional gauge coupling. The masses of the KK
gauge bosons $m_{i,j}$ are given in GeV units.
$M$ is the bulk gauge boson mass and $k^\prime = k/r_z$.
The non-zero values for $\frac{M}{k^\prime}$ are indicated in
the table. $\frac{1}{r_z}= 7\times 10^{17}$ GeV,
$k$ = 0.25 and $c$ = 11.52. The lowest lying mode $m_{1,1}$, which corresponds to $W$ or $Z$ boson, is not included in the table.}
\label{T:t1}
\end{table}
Now, in order to compute the gauge couplings the unknown parameters
that need to be fixed are $k$, $c$, $r_z$ and the bulk mass of the gauge fields
$M$. The non-zero value for the bulk gauge mass $M$ is around the Planck scale.
We can determine the remaining parameters by making the following demands:
(a) the lowest non-zero mass of the KK tower of the bulk gauge boson should be
identified with either $W$ or $Z$ boson mass, (b) the suppression $f$ of
eq. (\ref{E:supp}) should be $\sim 10^{-16}$ and (c) the hierarchy
between the moduli $R_y$ and $r_z$ should not be too large. The expression for the
KK gauge boson mass is given in eq. (\ref{E:xn}). For the lowest non-zero
KK gauge boson mass which is identifie as $m_{1,1}$ the root $x_{\nu_n}$ would be ${\cal O}(1)$. The factor $\exp(c\pi)
\cosh(k\pi)$ in this equation, which is the inverse of $f$,
should be $\sim 10^{16}$. By demanding that the lowest
KK mode $m_{1,1}$ has mass of $\sim 100$ GeV, from eq. (\ref{E:xn}) we can naively
estimate that $k^\prime\sim 10^{17}$ GeV. Since we would argue that
$k\sim 0.1$, a consistent value for the scale $r_z$ is
$\frac{1}{r_z}=7\times 10^{17}$ GeV, which is about 14 times smaller than
the Planck scale. The parameters $k$ and $c$ can
be determined from the fact that we should not get large hierarchy between
the moduli $R_y$ and $r_z$ and also we should get the desired suppression
of $f\sim 10^{-16}$ on the standard model brane. We have estimated that for
$k$ = 0.25 and $c$ = 11.52, the ratio between the moduli is $\frac{R_y}{r_z}$ = 61,
which is not unacceptably large, and also the suppression $f$
came out to be $1.45\times 10^{-16}$. For these particular values of
$k$, $c$ and $1/r_z$ we have given the gauge
couplings and the corresponding masses of the excited KK gauge fields
in Table \ref{T:t1}. In this table the gauge couplings $g^{(i,j)}$,
where $i,j$ are integers, of the standard
model fermion are given as a fraction of the 4-dimensional coupling
$g_0$. In the case of $\frac{M}{k^\prime}$ = 0.5 or = 1.0,
we have found that the lowest non-zero mode has mass of about 95 GeV.
Hence this mode can be identified with the $W$ or $Z$ gauge boson. Since
we have got the right amount of $W,Z$ boson masses for the above
described values of $k,c,\frac{1}{r_z}$, we use the same set of
values to get the gauge couplings and KK gauge boson masses in the
case where the bulk mass $M$ is zero. In this case we have found
that the lowest mode $m_{0,0}$ has zero mass which can be identified with
the photon state. The non-zero KK masses of the photon field and
their corresponding gauge coupling values are given in Table \ref{T:t2}.
\begin{table}
\begin{center}
\begin{tabular}{||c|c|c||} \hline
\multicolumn{3}{||c||}{$\frac{M}{k^\prime}=0$}\\ \hline
$m_{1,1}$ = 93.15 & $m_{1,2}$ = 142.0 & $m_{1,3}$ = 192.38 \\
$\tilde{g}^{1,1}$ = 0.0168 & $\tilde{g}^{1,2}$ = 0.0019 &
$\tilde{g}^{1,3}$ = 0.0028 \\\hline
$m_{2,1}$ = 178.45 & $m_{2,2}$ = 235.91 & $m_{2,3}$ = 293.12 \\
$\tilde{g}^{2,1}$ = 0.0128 & $\tilde{g}^{2,2}$ = 0.0015 &
$\tilde{g}^{2,3}$ = 0.0023 \\\hline
$m_{3,1}$ = 259.96 & $m_{3,2}$ = 320.71 & $m_{3,3}$ = 381.21 \\
$\tilde{g}^{3,1}$ = 0.0107 & $\tilde{g}^{3,2}$ = 0.0012 &
$\tilde{g}^{3,3}$ = 0.0020 \\\hline
\end{tabular}
\end{center}
\caption{The gauge couplings $g^{(0,0)(i,j)}$ of the standard model fermions
are given in the form $\tilde{g}^{i,j}=\frac{g^{(0,0)(i,j)}}{g_0}$, where
$g_0$ is the effective 4-dimensional gauge coupling. The masses of the KK
gauge bosons $m_{i,j}$ are given in GeV units.
$M$ is the bulk gauge boson mass, which is taken to be zero and $k^\prime = k/r_z$.
The values of $k$, $c$ and $r_z$ in this case are same as that in Table \ref{T:t1}. The lowest
lying mode $m_{0,0}$, which corresponds to photon, is not included in the table.}
\label{T:t2}
\end{table}

As stated in Sec. \ref{S:intro} that the 5-dimensional RS model
suffers from the precision electroweak tests due to the fact that
the first excited KK gague boson has coupling larger than one with
the standard model field. To parameterize the precision electroweak
constraints in extra dimensional models the following quantity
has been defined \cite{DHR,RW}.
\begin{equation}
V=\sum_n\left(\frac{g_n}{g_0}\frac{m_W}{M_n}\right)^2.
\label{E:peV}
\end{equation}
Here, $m_W$ is the mass of the $W$ gauge boson and $M_n$ is
the higher KK gauge boson mass.
The summation on $n$ in the above equation is over all the higher
KK gauge masses $M_n$ with corresponding gauge couplings $g_n$.
In our 6-dimensional model the index $n$ would be replaced by
a pair of integers and we should sum over all non-zero higher order modes.
It has been shown that by fitting to the precision electroweak
observables the quantity $V$ should satisfy the condition:
$V <$ 0.0013 at 95$\%$ confidence level \cite{DHR}. It can be easily checked
that this bound can be respected by the gauge couplings and
the KK masses of the tables \ref{T:t1} and \ref{T:t2}.
From both these tables, we can notice that the
gauge couplings are decreasing with increasing the KK gauge masses
for a particular value of $\frac{M}{k^\prime}$. Hence, in the summation
of eq. (\ref{E:peV}) only the first few higher KK modes are relevant.
We have checked that for $\frac{M}{k^\prime}$ = 0.5, $V$ has come out
to be about $5\times 10^{-5}$. In the case of photon where bulk mass
is zero the value of $V$ is found out to be about 0.000257. From these
results we can conclude that precision electroweak tests can be
satisfied in the 6-dimensional doubly warped model without
introducing too much hierarchy in the moduli $R_y$ and $r_z$.

At the Tevatron the higher KK gauge bosons have been searched
in the channel $P\bar{P}\to X\to e^+e^-$ and a
limit of $M_T>$ 700 GeV on the heavy vector gauge boson ($X$) mass has
been put-in \cite{Teva}.
In our case the gauge couplings of the higher KK modes have
been reduced from the 4-dimensional coupling $g_0$ by some
factors which are given in tables \ref{T:t1} and \ref{T:t2}.
Hence, in our case, the Tevatron bounds for the non-zero KK mode
masses should be greater than 700$\times\tilde{g}^{i,j}$ GeV.
As an example, the KK mode of 93.15 GeV mass of the Table \ref{T:t2}
has the gauge coupling ratio of 0.0168. Hence, the lower bound
from the Tevatron on this KK mode mass would be about 12 GeV,
which is much lower than our calculated value of 93.15 GeV.
Like wise, from each column of the tables \ref{T:t1} and \ref{T:t2} it
can be easily seen that the above mentioned Tevatron bounds
can be satisfied. So the 6-dimensional warped model is not only
free from the precision electroweak constraints but also from
the Tevatron limits.

\section{Conclusions}
The extra dimensional phenomenological models in a warped geometry encounters
problems in putting the Higgs and the gauge
fields in the bulk. It was shown that it is impossible to
construct proper $W$ and $Z$ boson masses on the brane from the KK modes of a
non-Abelian bulk gauge field through spontaneous symmetry breaking
in the bulk. Also proper coupling and masses for the first KK excitation of
a massless bulk gauge field consistent with
electroweak precision test as well as Fermilab Tevatron mass bound is hard to obtain
without changing the bulk parameter of the theory
from their desired values. In this work we have shown that it is possible to resolve both
these problems in a multiply warped geometry model
where there are more than one modulus. Considering a 6-dimensional model we have shown that
by setting one of the modulus approximately 
two orders smaller than the Planck scale, we can have the the mass for the lowest lying mode of the
bulk gauge field ( with bulk mass $\sim M_P$,
acquired through a spontaneous symmetry breaking in the bulk) on the TeV-brane 
to be of the order of $100$ GeV which therefore may be identified with the $W$, $Z$ boson mass.
Moreover, such a choice for the moduli which does
not contradict the main spirit of the RS model lowers the coupling of the first
KK mode excitation of a massless bulk gauge field so that
it can escape the electroweak precision test. We have determined the KK mode masses
as well as their couplings for different choices of
the parameter of the theory namely the ratio of the bulk mass and the bulk
cosmological constant. In the entire analysis the value of the
warp factor is maintained at $10^{-16}$ so that the resolution of the gauge hierarchy problem,
the main objective of these models can be achieved. These findings can be easily extended to models with even larger
number of warped extra dimensions \cite{cs}. One would then arrive at similar conclusions with a lesser hierarchy among different
moduli. We can therefore conclude that
a consistent description of bulk Higgs and gauge field with spontaneous symmetry breaking
in the bulk can be obtained in a warped geometry model
if the RS model in 5-dimensions is generalized to six or higher dimensions with more than one moduli. The phenomenology of these models 
therefore becomes an interesting area of study for the forthcoming collider experiments.

\end{document}